\documentclass[aps,preprint,showpacs,showkeys,nofootinbib]{revtex4}

\usepackage{ulem}
\usepackage{graphicx}  %
\usepackage{amsmath}
\usepackage{bm}  %

\newcommand{\bea}{\begin{eqnarray}}
\newcommand{\eea}{\end{eqnarray}}
\newcommand{\beq}{\begin{equation}}
\newcommand{\eeq}{\end{equation}}
\newcommand{\bqa}{\begin{eqnarray}}
\newcommand{\eqa}{\end{eqnarray}}

\def\mqo2{{\!\!\!}}

\begin{document}

\title{
How the $\bm{Z_c(3900)}$ Reveals the Spectra \\
of Quarkonium Hybrid and Tetraquark Mesons}
\author{Eric Braaten}
\affiliation{Department of Physics,
         The Ohio State University, Columbus, OH\ 43210, USA}
\date{\today}

\begin{abstract}
Flavor-exotic tetraquark mesons have recently been observed in the 
heavy-quark pair sectors of QCD, including two isospin multiplets 
in the $b \bar b$ sector, $Z_b(10610)$ and $Z_b(10650)$,
and one isospin multiplet in the $c \bar c$ sector, $Z_c(3900)$.
We identify $Z_b$ and $Z_c$ as tetraquark mesons 
that are analogs of quarkonium hybrids with the gluon field
replaced by an isospin-1 excitation of the light-quark fields.
Given the identification of $Y(4260)$ and $Z_c(3900)$ as
a ground-state charmonium hybrid and tetraquark, respectively,
lattice QCD calculations of the charmonium spectrum
can be used to estimate the masses of the lowest four 
spin-symmetry multiplets of charmonium hybrids and tetraquarks.
The $Z_b(10610)$ and $Z_b(10650)$ can be assigned to 
excited-state multiplets of bottomonium tetraquarks,
resulting in estimates of the masses of the ground-state  
multiplets of bottomonium hybrids and tetraquarks.
 \end{abstract}

\smallskip
\maketitle

One of the most basic problems of {\it Quantum Chromodynamics} (QCD)
is to identify all the clusters of quarks, antiquarks, and gluons
that are sufficiently bound by QCD interactions 
that they are either stable particles or sufficiently long-lived
to be observed as resonances.
The simplest bound clusters are {\it baryons}, 
which consist of three quarks ($qqq$), and ordinary {\it mesons}, 
which consist of a single quark and antiquark ($q \bar q$).  
The Review of Particle Properties
lists dozens of baryons that are well established,
hundreds of mesons that are well established,
and many more resonances that are less well established \cite{Beringer:1900zz}.  
The most general meson allowed by QCD is a cluster 
whose constituents consist of equally many quarks and antiquarks 
and possibly also gluons ($g$).  
An {\it exotic meson} has constituents that are not $q \bar q$.  
Two of the simplest types of exotic mesons are {\it hybrids}
 ($q \bar q g$) and {\it tetraquarks}  ($q q \bar q\bar q$).  
A {\it manifestly exotic meson} has quantum numbers that are
incompatible with $q \bar q$.  
A meson is spin-exotic if its $J^{PC}$ quantum numbers 
are in the sequence $\bm{0}^{--}$, $\bm{0}^{+-}$, $\bm{1}^{-+}$, 
$\bm{2}^{+-}$, $\bm{3}^{-+}$, \ldots.
(Here and below, a bold-face $\bm{J}$ 
indicates exotic quantum numbers.)
A meson is flavor-exotic if its flavor quantum numbers 
are incompatible with $q \bar q$. 
Until recently, no manifestly exotic mesons had been definitively identified.

The first definite discoveries of manifestly exotic mesons 
have come only in the last couple of years.  
Surprisingly, they have come in sectors of QCD 
that were believed to be the best understood: 
namely, the sectors containing a heavy quark and antiquark ($Q \bar Q$), 
where $Q$ can be a {\it charm quark} ($c$) 
or a {\it bottom quark} ($b$).  
In October 2011, the Belle collaboration announced the discovery 
of flavor-exotic tetraquark mesons $Z_b^+(10610)$ and $Z_b^+(10650)$ 
whose constituents are $b \bar b u \bar d$ \cite{Belle:2011aa}. 
(Their energies in MeV are given in parentheses).
Their widths are about 10~MeV and 20~MeV, respectively \cite{Belle:2011aa}.  
They were discovered through their decays into  
bottomonium and a pion:
$\Upsilon(nS)\, \pi^+$, $n=1,2,3$, and $h_b(nP)\, \pi^+$, $n = 1,2$.
The neutral member $Z_b^0(10610)$ of one of the isospin multiplets 
$(Z_b^-,Z_b^0, Z_b^+)$ has also been observed \cite{Adachi:2012im}.
These states have isospin and $G$-parity $I^G = 1^+$ 
and their preferred spin and parity are $J^P = 1^+$ \cite{Collaboration:2011gja}.  
In March 2013, the BESIII collaboration 
announced the discovery of a flavor-exotic tetraquark meson $Z_c^+(3900)$ 
whose constituents are $c \bar c u \bar d$ \cite{Ablikim:2013mio}.  
It was discovered through its decay into $J/\psi\, \pi^+$.
The existence of the $Z_c^+(3900)$ was confirmed 
by the Belle collaboration \cite{Liu:2013dau}
and by an analysis of data from the CLEOc collaboration \cite{Xiao:2013iha}.  
The latter analysis provided evidence for the neutral 
member $Z_c^0(3900)$ of the $(Z_c^-,Z_c^0, Z_c^+)$ multiplet.
The measurements in Refs.~\cite{Ablikim:2013mio,Liu:2013dau}
determine the mass of the $Z_c(3900)$ to be $3897 \pm 5$~MeV
and its width to be $51 \pm 18$~MeV.  

The constituents of the tetraquark mesons $Z_b$ and $Z_c$ 
are clearly revealed by their decay products.  
However the structure of these mesons is an open question.  
The possibilities that were proposed for $Z_c$ 
in the first two weeks after the announcement of its discovery 
include a cusp in $D^* \bar D$ scattering \cite{Wang:2013cya}, 
a charm-meson molecule consisting of $D^*$ and $\bar D$ 
\cite{Guo:2013sya,Chen:2013wca,Karliner:2013dqa,
Voloshin:2013dpa,Faccini:2013lda,Mahajan:2013qja,Cui:2013yva}, 
a tetraquark consisting of $c u$ and $\bar c \bar d$ diquarks 
\cite{Faccini:2013lda,Mahajan:2013qja},
and hadrocharmonium consisting of $u \bar d$ 
bound to a color-singlet $c \bar c$ core \cite{Voloshin:2013dpa,Mahajan:2013qja}.  
Many more possibilities for $Z_b$ have been proposed.  
These models are all inherently phenomenological,
making little direct contact with QCD.  
In particular, they are not easily verified through
nonperturbative calculations using lattice QCD.  
In this Letter, I point out compelling candidates for the $Z_b$ and $Z_c$ 
that can be confirmed using lattice gauge theory.  
I propose that they are analogs of quarkonium hybrids 
with the excitation of the gluon field
replaced by an isospin-1 excitation of the light-quark fields.

We begin by discussing quarkonium hybrids.  
Their existence in QCD without light quarks was demonstrated convincingly 
by Juge, Kuti, and Morningstar using the Born-Oppenheimer approximation 
in which the $Q$ and $\bar Q$ move slowly in response to the gluon field 
produced by static external sources \cite{Juge:1999ie}.  
They used lattice QCD without quarks to calculate the Born-Oppenheimer potentials
defined by the energy of the gluon field in the presence of a
color-triplet source and a color-antitriplet source separated by a distance $R$.
Ordinary quarkonia correspond to bound states 
of $Q$ and $\bar Q$ in the Born-Oppenheimer 
potential for the ground state of the gluon field,
which is labelled $\Sigma_g^+$.  
The bound states in this potential form spin-symmetry multiplets:  
$S$-wave multiplets with $J^{PC}$ quantum numbers 
$\{ 0^{-+},1^{--} \}$,
$P$-wave multiplets $\{ 1^{+-},(0,1,2)^{++} \}$,  
$D$-wave multiplets $\{ 2^{+-},(1,2,3)^{++} \}$, etc. 
Quarkonium hybrids correspond to bound states 
of $Q$ and $\bar Q$ in Born-Oppenheimer 
potentials for excited states of the gluon field.  
The lowest excited Born-Oppenheimer potentials 
are labeled $\Pi_u$ and  $\Sigma_u^-$.
The $\Pi_u$ potential is lower for $R>0$
and it has a minimum near 0.3~fm \cite{Juge:1999ie}.
The lowest bound states in the $\Pi_u$ potential
form a $P$-wave supermultiplet consisting of two spin-symmetry
multiplets, $H_0 = \{ 1^{--}, (0,\bm{1},2)^{-+} \}$
and $H_1 = \{ 1^{++}, (\bm{0},1,\bm{2})^{+-} \}$.
The degeneracy between these two multiplets is broken by the 
coupling between the orbital angular momentum of the heavy quarks 
and the angular momentum of the gluon field.
The spin-symmetry multiplets with the next lowest energies 
include $H_2 = \{ 0^{++}, 1^{+-} \}$,
which are $S$-wave bound states 
in the $\Sigma_u^-$ potential,
and $\{ 1^{--}, (0,\bm{1},2)^{-+} \}$,
which are radially excited $P$-wave bound states
in the $\Pi_u$ potential.
Juge, Kuti, and Morningstar used lattice NRQCD 
with bottom quarks but without light quarks 
to calculate the energies of the lowest bottomonium hybrids \cite{Juge:1999ie}.

For QCD with light quarks, the calculation of excited 
Born-Oppenheimer potentials for a heavy quark and antiquark 
is complicated by instabilities \cite{Bali:2000gf}.
For small separation $R$ of the $Q$ and $\bar Q$ sources, 
there is an instability with respect to 
a transition to the ground-state potential  
$\Sigma_g^+$ through the emission of two light mesons, such as $\pi^+ \pi^-$.
For large $R$, there can be an instability
with respect to decay into a pair of heavy mesons
corresponding to a light antiquark and quark 
localized near the $Q$ and $\bar Q$ sources, respectively.
If the transition rates for these instabilities are sufficiently small, 
it may still be possible 
to define Born-Oppenheimer potentials as almost stationary energy levels 
of light-quark and gluon fields in the 
presence of static $Q$ and $\bar Q$ sources. 
Quarkonium hybrids would be bound states in 
excited Born-Oppenheimer potentials with flavor-singlet quantum numbers.  
Quarkonium tetraquarks would be bound states in 
Born-Oppenheimer potentials with nontrivial flavor quantum numbers.  
The calculation of the tetraquark Born-Oppenheimer potentials
is a challenging problem for lattice QCD.

In the case of QCD with charm quarks and light quarks, 
the spectrum of charmonium hybrids
can be calculated directly using lattice gauge theory.  
Exploratory calculations of the spectrum of charmonium and 
charmonium hybrids above the charm-meson pair threshold 
have been carried out by Dudek, Richards, and Thomas \cite{Dudek:2009kk} 
and extended by the Hadron Spectrum Collaboration \cite{Liu:2012ze}.  
Since the calculations were carried out at a single lattice spacing 
and with $u$ and $d$ quark masses that correspond to a pion mass 
of about 400 MeV, 
the systematic errors could not be quantified.  
The lowest charmonium hybrids in the calculations of Ref.~\cite{Liu:2012ze}
form the multiplet $H_0$.  
Their quantum numbers coincide with those for an $S$-wave $c \bar c$ pair 
and an excitation of the gluon field with quantum numbers $1^{+-}$,
which corresponds to a constituent gluon in a $P$-wave state.  
There were also charmonium hybrid candidates corresponding 
to all members of a supermultiplet consisting of 
three spin-symmetry multiplets: 
$H_1$, $H_2$, and
$H_3 = \{ 2^{++}, (1,\bm{2},3)^{+-} \}$ \cite{Liu:2012ze}.
Their quantum numbers coincide with those for a $P$-wave $c \bar c$ pair 
and an excitation of the gluon field with quantum numbers $1^{+-}$.
This supermultiplet has been predicted to be the 
first excited multiplet for
hybrids containing light quarks \cite{Dudek:2011bn}. 

The results of Ref.~\cite{Liu:2012ze} are compatible with the 
identification of $Y(4260)$ as  the lowest $1^{--}$ charmonium hybrid. 
The $Y(4260)$ was discovered by the Babar collaboration 
in 2005 \cite{Aubert:2005rm} through its decay into $J/\psi\, \pi^+ \pi^-$.
It is a plausible candidate for a charmonium hybrid,
because it is produced so weakly in $e^+ e^-$ annihilation that
the peak of the resonance is near a local minimum of the  
hadronic cross section.
The small production rate in $e^+ e^-$ annihilation 
is a consequence of the small wavefunction for $c \bar c$ at the origin.
Annihilation decays into light hadrons
are also suppressed by the small wavefunction for $c \bar c$ at the origin.
One model-independent prediction for decays of a hybrid
is that decay into a pair of $S$-wave mesons 
is suppressed \cite{Kou:2005gt,Close:2005iz}.
The dominant decays of charmonium hybrids are therefore
expected to be into an $S$-wave and $P$-wave charm-meson pair, provided
these states are kinematically accessible.
The center of the $Y(4260)$ resonance is about 20~MeV
below the threshold for $D_1 \bar D$ and about 80~MeV above
the threshold for $D_0^* \bar D$, but the latter decay mode
is suppressed by a $D$-wave coupling.
Transition decays into charmonium plus light hadrons 
are also expected to have some suppression from 
the small overlap between the $c \bar c$ wavefunctions
for charmonium hybrids and charmonium.
Until the discovery of $Z_c$, the only decay modes 
of $Y(4260)$ that were observed were the discovery mode
$J/\psi\, \pi^+ \pi^-$ \cite{Aubert:2005rm} and the two additional
hadronic transition modes
$J/\psi\, \pi^0 \pi^0$ and $J/\psi\, K^+ K^-$ \cite{He:2006kg}.
The unexpectedly large branching fraction for two-pion 
transitions to $J/\psi$ was an obstacle to the definitive
identification of the $Y(4260)$ as a charmonium hybrid. 
The discovery of the $Z_c(3900)$ has removed that obstacle
by providing the new decay mode into $Z_c \pi$,
which contributes to $J/\psi\, \pi \pi$ through the 
subsequent decay $Z_c \to J/\psi\, \pi$.

Having identified the $Y(4260)$ as the lowest $1^{--}$ charmonium hybrid,
we can use the results of Ref.~\cite{Liu:2012ze} 
for the splittings between $c \bar c$ mesons
to estimate the masses of other charmonium hybrids. 
The other members of the lowest charmonium hybrid multiplet $H_0$
have quantum numbers $(0,\bm{1},2)^{-+}$ and their masses
are estimated to be $4173 \pm 21$, $4195 \pm 23$, and $4312 \pm 24$~MeV, 
respectively. (The errors are statistical uncertainties only.
They do not include the systematic errors associated 
with the extrapolation to zero lattice spacing 
or to the small physical masses of the $u$ and $d$ quarks.)
The centers of gravity of the multiplets $H_1$, $H_2$, and $H_3$
are estimated to be 4361, 4454, and 4495, respectively.

We now turn to the isospin-1 charmonium tetraquark $Z_c(3900)$.
I propose that $Z_c$ is the analog of a charmonium hybrid 
with the excitation of the gluon field
replaced by an isospin-1 excitation of the light-quark fields.
Recall that the $Y(4260)$ can be interpreted as  
a $0^{-+}$ $c \bar c$ pair plus a $1^{+-}$ gluon.  
I assume that $Z_c$ is a $0^{-+}$ $c \bar c$ pair 
plus a $q \bar q$ with isospin 1  
and that the decay $Y \to Z_c \, \pi$ proceeds
through a transition of $g$ 
into $q \bar q$ by pion emission,
with the $c \bar c$ pair acting essentially as a spectator.
The pion has $I^G(J^P) = 1^-(0^-)$,
and its Goldstone nature requires it to be emitted in a $P$-wave state.
This implies that the $q \bar q$ in the $Z_c$
has $I^G = 1^+$ and $J^P = 0^+$, $1^+$, or $2^+$.
Equivalently, the $q \bar q$ in the neutral 
tetraquark $Z_c^0$ has $J^{PC} = 0^{+-}$, $1^{+-}$, or $2^{+-}$.
Given that the lowest-energy gluon excitation in a 
quarkonium hybrid is $1^{+-}$,
it is plausible that the lowest-energy light-quark excitation in a 
neutral quarkonium tetraquark is also $1^{+-}$.
I will assume that this is indeed the case, so
the $q \bar q$ in the $Z_c$ is $I^G(J^P) = 1^+(1^+)$.
Since the $Z_c$ also contains a $0^{-+}$ $c \bar c$ pair,
it must be $1^+(1^-)$.
It has negative parity, in contrast to most previous theoretical 
interpretations of the $Z_c$, which assumed that its parity is positive
\cite{Wang:2013cya,Guo:2013sya,Chen:2013wca,Karliner:2013dqa,Faccini:2013lda,
Voloshin:2013dpa,Mahajan:2013qja,Cui:2013yva}.  
The quantum numbers $J^P = 1^-$ allow decays of $Z_c$ into $D \bar D$ 
and $D^* \bar D$, but decays into these $S$-wave charm meson pairs  
are suppressed for the same reason they are suppressed
for charmonium hybrids \cite{Kou:2005gt,Close:2005iz}.
Annihilation decays into light hadrons are suppressed
by the small wavefunction for $c \bar c$ at the origin,
again like for charmonium hybrids.
The dominant decays of $Z_c$
should therefore be hadronic transitions to charmonium. 
In addition to the discovery decay mode $J/\psi\, \pi$, 
the other such 2-body decay modes are $\psi(2S)\, \pi$ and $\eta_c\, \rho$.

The other three members of the spin-symmetry multiplet $T_0$ of $Z_c$
are obtained by replacing the $0^{-+}$ $c \bar c$ pair 
by a $1^{--}$  $c \bar c$ pair.  
The ground-state tetraquark multiplet 
is therefore $T_0 = \{ 1^+(1^-), 1^-(0^-,\bm{1}^-,2^-) \}$.  
Assuming the mass splittings relative to the corresponding
charmonium hybrids are the same as the splitting between 
$Z_c(3900)$ and $Y(4260)$,
the estimates for the masses of the $(0,\bm{1},2)^-$ states 
are 3807, 3829, and 3946~MeV, respectively.
The spin-$J$ state can decay into $\chi_{cJ}\,  \pi$
and all three states can decay into $J/\psi\, \rho$, $\eta_c\, \pi$, and
$\eta_c(2S)\, \pi$.
The lowest excited isospin-1 charmonium tetraquarks 
form three spin-symmetry multiplets
analogous to those for charmonium hybrids 
in Ref.~\cite{Liu:2012ze}:
$T_1 = \{ 1^-(1^+), 1^+(\bm{0}^+,1^+,\bm{2}^+) \}$, 
$T_2 = \{ 1^-(0^+), 1^+(1^+) \}$,
and $T_3 = \{ 1^-(2^+), 1^+(1^+,\bm{2}^+,3^+,) \}$. 
Their centers of gravity are estimated to be near 
3995, 4088, and 4130~MeV, respectively.
It should be possible to calculate the masses of these charmonium 
tetraquarks using lattice QCD by the methods of Ref.~\cite{Liu:2012ze}.
Their spectrum is more sensitive 
to the extrapolation to the physical masses of the $u$ and $d$ quarks
than the spectrum of charmonium hybrids.

We now turn to the isospin-1 bottomonium tetraquarks 
$Z_b(10610)$ and $Z_b(10650)$.
Their quantum numbers are $I^G(J^P) = 1^+(1^+)$.  
Their measured masses are fortuitously close to the 
$B^* \bar B$ and $B^* \bar B^*$ thresholds, respectively.  
Since they have $S$-wave couplings to these thresholds, 
their resonant interactions with pairs of bottom mesons 
transform them into loosely-bound molecules \cite{Braaten:2003he}.  
The $Z_b(10610)$ has constituents $B^* \bar B$ or $B \bar B^*$ 
and the $Z_b(10650)$ has constituents $B^* \bar B^*$.  
The decay widths of these loosely-bound molecules 
are suppressed by the large mean separation of the constituents, 
scaling with their small binding energies $E_b$ 
as $E_b^{1/2}$ \cite{Braaten:2003he}.  
The phenomenological description of $Z_b(10610)$ and $Z_b(10650)$ 
as loosely-bound molecules has been fairly successful 
\cite{Bondar:2011ev,Cleven:2011gp,Mehen:2011yh,
Dong:2012hc,Li:2012uc,Ohkoda:2012rj,Cleven:2013sq}.

I propose that $Z_b(10610)$ and $Z_b(10650)$ both arise from
accidental fine tunings of $b \bar b$ tetraquarks to the 
$B^* \bar B$ and $B^* \bar B^*$ thresholds.
If not for the effects of $S$-wave rescattering of pairs of bottom mesons,
these tetraquarks would be bottomonium hybrids
with the excitation of the gluon field
replaced by an isospin-1 excitation of the light-quark fields.
The lowest energy bottomonium hybrids form the spin-symmetry multiplets
$H_0$,  $H_1$, and $H_2$ defined above.
Bottomonium tetraquarks that differ only by the replacement
of a $1^{+-}$ gluon excitation
by an isospin-1 $1^{+-}$ light-quark excitation form the multiplets
$T_0$,  $T_1$, and $T_2$ defined above.
The multiplets $T_1$ and $T_2$ include $1^+(1^+)$ states 
that can be identified with the $Z_b(10610)$ and $Z_b(10650)$.
The four isovector multiplets $1^-(1^+)$, $1^+(\bm{0}^+)$, 
$1^+(\bm{2}^+)$, and $1^-(0^+)$ should be nearby in energy,
with splittings less than the 46~MeV $B^*-B$ splitting.
This set of quantum numbers does not coincide with 
the sets of nearby states predicted by any of the molecular models
of the $Z_b$ states \cite{Bondar:2011ev,Cleven:2011gp,Mehen:2011yh,
Dong:2012hc,Li:2012uc,Ohkoda:2012rj,Cleven:2013sq}.
The proximity of $Z_b(10610)$ and $Z_b(10650)$  to the 
$B^* \bar B$ and $B^* \bar B^*$ thresholds is regarded as fortuitous.
Such accidental fine tunings may at first seem very unlikely.
However the bottom mesons $B$ and $B^*$ produce three pair thresholds
with a total splitting of 92~MeV.
The two multiplets that include $1^+(1^+)$ states 
contain a total of 6 states.  
Two of them have $S$-wave couplings to $D \bar D$,
three of them have $S$-wave couplings to $D^* \bar D$,
and all of them have $S$-wave couplings to $D^* \bar D^*$,
If both multiplets happen to be in the region of
the $B \bar B$, $B^* \bar B$, and  $B^* \bar B^*$ thresholds,
it is not so unlikely that two of the 6 states happen to be much closer 
to a threshold than the $B^*-B$ splitting.
The remaining 4 states need not be near a threshold
to which they have an $S$-wave coupling, 
so they are not expected to have the same molecular character
as the $Z_b(10610)$ and $Z_b(10650)$. 

We can use our identification of  $Z_b(10610)$ and $Z_b(10650)$
as members of excited bottomonium tetraquark multiplets to estimate
the masses of the ground-state tetraquark and hybrid multiplets.
The differences between the centers of gravity of the multiplets 
should be approximately the same for bottomonium and charmonium,
while the spin splittings within multiplets should be smaller 
for bottomonium by about a factor of 3.
The estimates for the center of gravity of the 
ground-state tetraquark multiplet $T_0$
using the masses of $Z_b(10610)$ and $Z_b(10650)$ as inputs are
10457 and 10519~MeV, respectively.
The estimates for the center of gravity of the 
ground-state hybrid multiplet $H_0$ are
10823 and 10884~MeV, respectively.
A candidate for the $1^{--}$ member of this multiplet is the
$Y_b(10888)$ observed by the Belle collaboration as a resonance 
in $e^+ e^-$ annihilation into $\Upsilon(nS)\, \pi^+ \pi^-$
that overlaps with the $\Upsilon (5S)$ \cite{Adachi:2008pu}.
In addition to the observed decays of $\Upsilon (5S)$ and/or
$Y_b(10888)$ into $Z_b(10610)\, \pi$ and $Z_b(10650)\, \pi$,
there should also be a substantial decay mode into 
$Z_b\, \pi$, where $Z_b$ is the $1^+(1^-)$ member of the 
ground-state tetraquark multiplet $T_0$.
Using the mass of $Y_b(10888)$ as input,
we can estimate the masses for the three other
ground-state hybrids in $H_0$
by dividing the mass splittings for ground-state charmonium 
hybrids in  Ref.~\cite{Liu:2012ze} by a factor of 3.  
The resulting estimates for the masses of the $1^+(0,\bm{1},2)^+$ 
hybrids are 10858, 10866, and 10905~MeV, respectively.

We have assumed that hadronic transitions between a quarkonium hybrid
and the corresponding tetraquark proceed by a transition 
between the excitations of the light-quark and gluon fields,
with the $Q$ and $\bar Q$ behaving as spectators.
This implies that the  $\Pi_u$ Born-Oppenheimer potentials
for hybrids and tetraquarks are the same, except for an offset that
is approximately equal to the mass splitting betwen
$Y(4260)$ and $Z_c(3900)$.
The pair of sources for the $Q$ and $\bar Q$ reduce 
in the limit $R \to 0$ to a single color-octet source.
Because of the local $SU(3)$ gauge symmetry of QCD,
it is only in the limit $R \to 0$ that the color state of the 
$Q \bar Q$ pair can be unambiguously identified as color-octet.
If we nevertheless interpret the color state of the 
$Q \bar Q$ pair  for $R>0$ as color-octet by continuity,
a quarkonium tetraquark can be interpreted as an isospin-1 
color-octet $q \bar q$ pair bound to a color-octet $Q \bar Q$ pair.
A relatively simple lattice QCD calculation that would shed light 
on quarkonium hybrids and tetraquarks 
would be the spectrum of light-quark and gluon fields
in the presence of a static color-octet source.
Only recently have lattice gauge configurations become available with 
sufficiently light $u$ and $d$ quarks 
to answer this question definitively.
My expectation is that the energy of the lowest flavor-singlet state 
is higher than that of the lowest isospin-1 state
by an amount comparable to the splitting between 
$Y(4260)$ and $Z_c(3900)$.

Before the discovery of $Z_c$, several unconfirmed
isospin-1 charmonium tetraquarks were observed
in $B$ meson decays.  
The $Z^+(4430)$ was observed through its decay into 
$\psi(2S)\, \pi^+$ \cite{Choi:2007wga}. 
Its mass is a couple hundred MeV too high to be 
assigned to the multiplet $T_3$.
The $Z^+(4050)$ and $Z^+(4250)$ were observed through their decays
into $\chi_{c1}\, \pi^+$ \cite{Mizuk:2008me}. 
The mass of $Z(4050)$ is close to the estimated center of gravity 
of the $T_2$ multiplet.  The $\chi_{c1}\, \pi$ decay mode is compatible 
with the $1^-(0^+)$ member of that multiplet.

More than a dozen neutral $XYZ$ charmonium states 
have been observed \cite{Brambilla:2010cs}.
Those that decay into pairs of $S$-wave states can be excluded 
as candidates for charmonium hybrids or tetraquarks.
None of the remaining $XYZ$ states are
candidates for a neutral isospin-1 tetraquark.
The $X(4350)$, which decays into $J/\psi\, \phi$ 
and $\gamma \gamma$
and has quantum numbers $0^{P+}$ or $2^{P+}$ 
\cite{Shen:2009vs}, is a good candidate
for the $0^{++}$ member of the $H_1$ hybrid multiplet.
The $X(4140)$ and $Y(4274)$, which have been observed in 
$J/\psi\, \phi$ \cite{Yi:2010aa}, are candidates for the $H_0$ hybrid multiplet.
There are several $Y$ states with quantum numbers $1^{--}$ 
that cannot be accomodated by any of the hybrid or tetraquark 
multiplets discussed above.
They could be isospin-0 quarkonium tetraquarks  
related to the isospin-1 tetraquarks by $SU(3)$ flavor symmetry.
The $X(3915)$, which decays into $J/\psi\, \omega$  and $\gamma \gamma$ 
and has quantum numbers $0^{P+}$ \cite{Lees:2012xs},
is a good candidate for the $0^{-+}$ state of that multiplet.  
An alternative identification of $X(3915)$ as the $P$-wave charmonium state
$\chi_{c0}(2P)$ is disfavored by its not having been observed 
in the decay mode $D \bar D$ \cite{Guo:2012tv}.
Flavor symmetry suggests that there are also strange
charmonium tetraquarks with energies a couple  hundred MeV higher 
than their nonstrange counterparts.

In summary, the flavor-exotic mesons $Z_c$ and $Z_b$
have been identified as quarkonium tetraquarks related 
to quarkonium hybrids  by replacing the gluon excitation 
by an isospin-1 light-quark excitation.
Lattice QCD calculations of the 
charmonium spectrum were used to estimate the masses 
of the lowest spin-symmetry multiplets of
quarkonium hybrids and tetraquarks.
Many of the remaining $XYZ$ mesons fit naturally 
into one of those hybrid multiplets or into isospin-0 
tetraquark multiplets related to the isospin-1 multiplets
by $SU(3)$ flavor symmetry.
A rich spectrum of additional quarkonium hybrids and tetraquarks 
is awaiting discovery.

\begin{acknowledgments}
This research was supported in part by the Department of Energy 
under grant DE-FG02-91-ER40690.
\end{acknowledgments}


\begin{thebibliography}{99}

\bibitem{Beringer:1900zz} 
  J.~Beringer {\it et al.}  [Particle Data Group Collaboration],
  Phys.\ Rev.\ D {\bf 86}, 010001 (2012).

\bibitem{Belle:2011aa} 
  A.~Bondar {\it et al.}  [Belle Collaboration],
  Phys.\ Rev.\ Lett.\  {\bf 108}, 122001 (2012)
  [arXiv:1110.2251].
  
\bibitem{Adachi:2012im} 
  I.~Adachi {\it et al.}  [Belle Collaboration],
  arXiv:1207.4345.
  
\bibitem{Collaboration:2011gja} 
  I.~Adachi [Belle Collaboration],
  arXiv:1105.4583.
  
\bibitem{Ablikim:2013mio} 
  M.~Ablikim {\it et al.}  [BESIII Collaboration],
  arXiv:1303.5949.
  
\bibitem{Liu:2013dau} 
  Z.Q.~Liu {\it et al.}  [Belle Collaboration],
  arXiv:1304.0121.
  
\bibitem{Xiao:2013iha} 
  T.~Xiao, S.~Dobbs, A.~Tomaradze and K.K.~Seth,
  arXiv:1304.3036.
  
\bibitem{Wang:2013cya} 
  Q.~Wang, C.~Hanhart and Q.~Zhao,
  arXiv:1303.6355.

\bibitem{Guo:2013sya} 
  F.-K.~Guo, C.~Hidalgo-Duque, J.~Nieves and M.P.~Valderrama,
  arXiv:1303.6608.  

\bibitem{Chen:2013wca} 
  D.-Y.~Chen, X.~Liu and T.~Matsuki,
  arXiv:1303.6842.

\bibitem{Faccini:2013lda} 
  R.~Faccini, L.~Maiani, F.~Piccinini, A.~Pilloni, A.D.~Polosa and V.~Riquer,
  arXiv:1303.6857.
  
\bibitem{Karliner:2013dqa} 
  M.~Karliner and S.~Nussinov,
  arXiv:1304.0345.

\bibitem{Voloshin:2013dpa} 
  M.B.~Voloshin,
  arXiv:1304.0380.
  
\bibitem{Mahajan:2013qja} 
  N.~Mahajan,
  arXiv:1304.1301.
  
\bibitem{Cui:2013yva} 
  C.-Y.~Cui, Y.-L.~Liu, W.-B.~Chen and M.-Q.~Huang,
  arXiv:1304.1850.
  
\bibitem{Juge:1999ie} 
  K.J.~Juge, J.~Kuti and C.J.~Morningstar,
  Phys.\ Rev.\ Lett.\  {\bf 82}, 4400 (1999)
  [hep-ph/9902336].
  
\bibitem{Bali:2000gf} 
  G.S.~Bali,
  Phys.\ Rept.\  {\bf 343}, 1 (2001)
  [hep-ph/0001312].
  
\bibitem{Dudek:2009kk} 
  J.J.~Dudek, R.G.~Edwards and C.E.~Thomas,
  Phys.\ Rev.\ D {\bf 79}, 094504 (2009)
  [arXiv:0902.2241].  
  
\bibitem{Liu:2012ze} 
  L.~Liu {\it et al.}  [Hadron Spectrum Collaboration],
  JHEP {\bf 1207}, 126 (2012)
  [arXiv:1204.5425].

\bibitem{Dudek:2011bn} 
  J.J.~Dudek,
  Phys.\ Rev.\ D {\bf 84}, 074023 (2011)
  [arXiv:1106.5515].
  
\bibitem{Aubert:2005rm} 
  B.~Aubert {\it et al.}  [BaBar Collaboration],
  Phys.\ Rev.\ Lett.\  {\bf 95}, 142001 (2005)
  [hep-ex/0506081].
  
\bibitem{Kou:2005gt} 
  E.~Kou and O.~Pene,
  Phys.\ Lett.\ B {\bf 631}, 164 (2005)
  [hep-ph/0507119].
  
\bibitem{Close:2005iz} 
  F.E.~Close and P.R.~Page,
  Phys.\ Lett.\ B {\bf 628}, 215 (2005)
  [hep-ph/0507199].
  
\bibitem{He:2006kg} 
  Q.~He {\it et al.}  [CLEO Collaboration],
  Phys.\ Rev.\ D {\bf 74}, 091104 (2006)
  [hep-ex/0611021].
  
\bibitem{Braaten:2003he} 
  E.~Braaten and M.~Kusunoki,
  Phys.\ Rev.\ D {\bf 69}, 074005 (2004)
  [hep-ph/0311147].
  
\bibitem{Bondar:2011ev} 
  A.E.~Bondar, A.~Garmash, A.I.~Milstein, R.~Mizuk and M.B.~Voloshin,
  Phys.\ Rev.\ D {\bf 84}, 054010 (2011)
  [arXiv:1105.4473].
  
\bibitem{Cleven:2011gp} 
  M.~Cleven, F.-K.~Guo, C.~Hanhart and Ulf-G.~Meissner,
  Eur.\ Phys.\ J.\ A {\bf 47}, 120 (2011)
  [arXiv:1107.0254].
  
\bibitem{Mehen:2011yh} 
  T.~Mehen and J.W.~Powell,
  Phys.\ Rev.\ D {\bf 84}, 114013 (2011)
  [arXiv:1109.3479].

\bibitem{Dong:2012hc} 
  Y.~Dong, A.~Faessler, T.~Gutsche and V.E.~Lyubovitskij,
  J.\ Phys.\ G {\bf 40}, 015002 (2013)
  [arXiv:1203.1894].

\bibitem{Li:2012uc} 
  X.~Li and M.B.~Voloshin,
  Phys.\ Rev.\ D {\bf 86}, 077502 (2012)
  [arXiv:1207.2425].

\bibitem{Ohkoda:2012rj} 
  S.~Ohkoda, Y.~Yamaguchi, S.~Yasui and A.~Hosaka,
  Phys.\ Rev.\ D {\bf 86}, 117502 (2012)
  [arXiv:1210.3170].
  
\bibitem{Cleven:2013sq} 
  M.~Cleven, Q.~Wang, F.-K.~Guo, C.~Hanhart, Ulf-G.~Meissner and Q.~Zhao,
  Phys.\ Rev.\ D {\bf 87}, 074006 (2013)
  [arXiv:1301.6461].

\bibitem{Adachi:2008pu} 
  I.~Adachi {\it et al.}  [Belle Collaboration],
  arXiv:0808.2445.  
  
\bibitem{Aubert:2009aq} 
  B.~Aubert {\it et al.}  [BaBar Collaboration],
  Phys.\ Rev.\ D {\bf 79}, 092001 (2009)
  [arXiv:0903.1597].
  
\bibitem{Brambilla:2010cs} 
  N.~Brambilla, S.~Eidelman, B.K.~Heltsley, R.~Vogt, G.T.~Bodwin, 
  E.~Eichten, A.D.~Frawley and A.B.~Meyer {\it et al.},
  Eur.\ Phys.\ J.\ C {\bf 71}, 1534 (2011)
  [arXiv:1010.5827].
  
\bibitem{Choi:2007wga} 
  S.K.~Choi {\it et al.}  [BELLE Collaboration],
  Phys.\ Rev.\ Lett.\  {\bf 100}, 142001 (2008)
  [arXiv:0708.1790].  
  
\bibitem{Mizuk:2008me} 
  R.~Mizuk {\it et al.}  [Belle Collaboration],
  Phys.\ Rev.\ D {\bf 78}, 072004 (2008)
  [arXiv:0806.4098].  

\bibitem{Shen:2009vs} 
  C.P.~Shen {\it et al.}  [Belle Collaboration],
  Phys.\ Rev.\ Lett.\  {\bf 104}, 112004 (2010)
  [arXiv:0912.2383].
  
\bibitem{Yi:2010aa} 
  K.~Yi [CDF Collaboration],
  PoS ICHEP {\bf 2010}, 182 (2010)
  [arXiv:1010.3470].
  
\bibitem{Lees:2012xs} 
  J.P.~Lees {\it et al.}  [BaBar Collaboration],
  Phys.\ Rev.\ D {\bf 86}, 072002 (2012)
  [arXiv:1207.2651].
  
\bibitem{Guo:2012tv} 
  F.-K.~Guo and Ulf-G.~Meissner,
  Phys.\ Rev.\ D {\bf 86}, 091501 (2012)
  [arXiv:1208.1134].
  
  \end{thebibliography}
\end{document}